\documentclass[useAMS,usenatbib]{mn2e}

\usepackage{graphicx}

\title[Radial velocity study of $\epsilon$ UMa]
{Radial velocity study of the CP star $\epsilon$ Ursae Majoris}

\author[N. A. Sokolov]{N. A. Sokolov$^{1,2}$
\thanks{E-mail: sokolov@gao.spb.ru}
\thanks{Based on observations collected with the ELODIE spectrograph
on the 193-cm telescope at the Observatoire de Haute-Provence (CNRS), France}\\
$^{1}$Central Astronomical Observatory at Pulkovo,
St. Petersburg 196140, Russia\\
$^{2}$Isaac Newton Institute of Chile, Branch at St. Petersburg}

\begin{document}

\date{Accepted 2007 November 16. Received 2007 November 16; in original form 2007 July 27}

\pagerange{\pageref{firstpage}--\pageref{lastpage}} \pubyear{2008}

\maketitle

\label{firstpage}

\begin{abstract}
In this Letter, the radial velocity variability of the chemically
peculiar star $\epsilon$~Ursae Majoris ($\epsilon$~UMa) from the sharp
cores of the hydrogen lines is investigated.
This study is based on the ELODIE archival data obtained
at different phases of the rotational cycle.
The star exhibits low-amplitude radial velocity variations with a
period of $P$~=~5.0887 d. The best Keplerian solution yields
an eccentricity $e$~=~0.503 and a minimum mass $\sim$14.7$M_{\rm Jup}$
on the hypothesis that the rotational axis of $\epsilon$~UMa is
perpendicular to the orbital plane.
This result indicate that the companion is
the brown-dwarf with the projected semi-amplitude variation of the
radial velocity $K_{\rm 2}$~=~135.9~km~${\rm s}^{\rm -1}$ and
the sine of inclination times semi-major axis
$a_{2}$$\sin$(i)~=~0.055~au.
\end{abstract}


\section{Introduction}

Epsilon Ursae Majoris ($\epsilon$ UMa, HD~112185, HR~4905) is
the brightest ($V$~=~1.77~mag) chemically peculiar (CP) star and
has been extensively studied during the last century.
\citet{b1} established a period of 5.$^{\rm d}$0887 d from variations in
the intensity of the Ca~II~K line and also noticed a periodic splitting
of some lines.
\citet{b2} subsequently reported doubling of lines of Cr~II, Fe~II, V~II
and other elements at certain phases.
Since the overall widths of the double lines are the same and not
all lines double, they ruled out the orbital motion and instead suggested
that the phenomenon is related to rotation of the star.
\citet{b3} measured a double wave light variation with the same
5.$^{\rm d}$0887 period. The star is brightest when the Ca~II~K line
intensity is near its minimum and other elements are near their maximum
strength.

Perhaps the most interesting and controversial aspect of $\epsilon$~UMa
is its the radial velocity variations.
\citet{b4} carried out the radial velocity analysis of the 62 brightest
northern CP stars for spectroscopic binaries.
They were used special observing and measuring techniques of the hydrogen
lines and did not find the radial velocity variations for $\epsilon$~UMa.
Not surprisingly, they could not to measure the radial velocity
variations because the mean internal error per spectrum of
1.2~km~${\rm s}^{-1}$ is near the limit at which the photographic technique
is useful.
On the other hand, \citet{b12} detected the duplicity for the system
$\epsilon$~UMa with a separation of 0.053 arcsec, using speckle interferometry
technique. Although, the authors noted that the observed separation
corresponds to the diffraction limit of the 2.5-m Isaac Newton telescope.

\citet{b5} measured the radial velocity variations of many lines in
the star and found sinusoidal variations in the radial velocities of Fe, Cr
and Ti lines with amplitudes of about 20~km~${\rm s}^{-1}$ and attributed this
to the existence of a spots of enhanced abundance of these elements.
Several surface abundance Doppler images of $\epsilon$~UMa have been produced
\citep{b6, b7, b8, b9, b10}.
Most of the published Doppler images relate to Fe, O, Ca, or Cr abundance.
Recently, \citet{b11} determined for $\epsilon$~UMa for the first time
the abundance distributions of Mn, Ti, Sr, and Mg.
Attempt to determine orbital motion using these spectral lines would be
impossible. However, the hydrogen lines do not show significant
rotational effect in their radial velocity. They would be preferable for
the measurement of binary motion if the radial velocity of the hydrogen lines
could be measured with precision.

In this Letter, we present the radial velocity measurement of the hydrogen
lines in the spectrum of CP star $\epsilon$~UMa. The ELODIE observations
of this star and data reduction are described in Sect.~2.
The orbital solution derived for the satellite candidate are presented
in Sect.~3. Section 4 reports discussion of our results for $\epsilon$~UMa
and conclusions are presented in Sect.~5.

\section{Observational material and data reduction}

\begin{table*}
 \centering
 \begin{minipage}{150mm}
 \caption{Measured radial velocities of $\epsilon$ UMa. All data are relative
 to the Solar system barycentre.}
  \begin{tabular}{@{}crccrrrrrr@{}}
  \hline
 File  name& Exposure time & MJD & Phase & \multicolumn{4}{c}{Radial velocity (km~${\rm s}^{-1}$)}& & \\
 & (s) &(2,450,000+)& & $H_{\delta}$ & $H_{\gamma}$ & $H_{\beta}$ & $H_{\alpha}$ &
 RV$_{Mean}$ & $\sigma_{Mean}$\\[5pt]
\hline
19960204 0022 & 168.73 & 0117.9814 & 0.821 & -9.56 & -9.19 & -9.93 & -9.47 & -9.54 & 0.31\\
19970218 0032 & 200.81 & 0497.9778 & 0.496 & -9.06 & -8.72 & -9.67 & -9.27 & -9.18 & 0.40\\
19970218 0033 & 200.37 & 0497.9819 & 0.497 & -8.90 & -8.77 & -9.51 & -9.19 & -9.09 & 0.33\\
20000124 0030 &  60.58 & 1568.1285 & 0.795 & -9.60 & -9.07 & -9.73 & -9.48 & -9.47 & 0.29\\
20000124 0031 &  90.57 & 1568.1440 & 0.798 & -9.50 & -9.16 & -9.71 & -9.63 & -9.50 & 0.24\\
20000126 0040 & 101.56 & 1570.1447 & 0.191 & -8.11 & -8.24 & -8.99 & -8.94 & -8.57 & 0.46\\
20000126 0041 &  50.58 & 1570.1499 & 0.192 & -8.29 & -8.28 & -9.07 & -8.99 & -8.66 & 0.43\\
20000519 0012 & 120.79 & 1683.9584 & 0.557 & -9.04 & -8.82 & -9.17 & -8.99 & -9.01 & 0.14\\
20000520 0009 & 120.76 & 1684.8706 & 0.737 & -9.56 & -9.02 & -9.32 & -9.13 & -9.26 & 0.24\\
20000521 0006 & 600.78 & 1685.8694 & 0.933 &-10.14 & -9.49 & -9.71 & -9.36 & -9.68 & 0.34\\
20000521 0007 & 600.44 & 1685.8784 & 0.935 & -9.85 & -9.44 & -9.63 & -9.44 & -9.59 & 0.20\\
20000606 0017 &  30.54 & 1701.8490 & 0.073 & -8.05 & -7.89 & -9.04 & -8.55 & -8.38 & 0.52\\
20000607 0007 &  30.27 & 1702.8632 & 0.272 & -8.44 & -8.33 & -8.73 & -8.44 & -8.48 & 0.17\\
20000608 0006 &  30.53 & 1703.8467 & 0.466 & -9.09 & -8.80 & -9.24 & -8.87 & -9.00 & 0.20\\
20000609 0003 &  30.97 & 1704.8443 & 0.662 & -9.34 & -8.64 & -9.35 & -8.68 & -9.00 & 0.40\\
20000615 1417 & 300.17 & 1710.8469 & 0.841 & -9.51 & -8.94 & -9.76 & -9.08 & -9.32 & 0.38\\
20000615 1420 &  50.47 & 1710.8505 & 0.842 & -9.46 & -8.74 & -9.79 & -9.30 & -9.32 & 0.44\\
20000615 1423 &  30.76 & 1710.8608 & 0.844 & -9.61 & -8.63 & -9.77 & -9.05 & -9.27 & 0.52\\
20000615 1425 &  30.26 & 1710.8634 & 0.844 & -9.68 & -8.84 & -9.67 & -9.17 & -9.34 & 0.41\\
20000615 1426 &  30.26 & 1710.8661 & 0.845 & -9.67 & -8.99 &-10.04 & -9.13 & -9.46 & 0.49\\
20000616 1418 &  50.92 & 1711.8591 & 0.040 & -8.35 & -7.88 & -9.11 & -8.85 & -8.55 & 0.55\\
20000616 1421 &  20.71 & 1711.8630 & 0.041 & -8.17 & -7.81 & -9.17 & -8.94 & -8.52 & 0.64\\
20000616 1424 &  20.65 & 1711.8670 & 0.042 & -8.13 & -8.00 & -9.33 & -8.94 & -8.60 & 0.64\\
20000618 0006 &  30.59 & 1713.8460 & 0.431 & -8.52 & -8.45 & -9.04 & -8.80 & -8.70 & 0.27\\
20000618 0007 &  30.71 & 1713.8506 & 0.432 & -8.52 & -8.40 & -8.76 & -8.74 & -8.61 & 0.17\\
20000618 0008 &  30.97 & 1713.8542 & 0.432 & -8.43 & -8.42 & -8.84 & -8.80 & -8.62 & 0.23\\
20000619 0024 &  30.49 & 1714.8368 & 0.625 & -8.85 & -8.47 & -9.47 & -8.77 & -8.89 & 0.42\\
20000619 0025 &  30.54 & 1714.8715 & 0.632 & -9.37 & -8.68 & -9.22 & -8.76 & -9.01 & 0.34\\
20000619 0026 &  30.04 & 1714.8742 & 0.633 & -9.23 & -8.96 & -9.17 & -8.83 & -9.05 & 0.19\\
\hline
\end{tabular}
\end{minipage}
\end{table*}

\begin{table}
 \centering
 \begin{minipage}{150mm}
 \caption{Best Keplerian orbital solution derived for $\epsilon$ UMa}
  \begin{tabular}{@{}llcc@{}}
  \hline
 Parameter & & Value & Error \cr
\hline
{\em P} (fixed) & (d) & 5.0887 &       \\
{\em T} (periastron) & (JD-2450000)   & 1743.932 & 0.161 \\
{\em e} &                &   0.503  & 0.063 \\
${V}_{\rm 0}$ & (km~${\rm s}^{-1}$) & -8.920 & 0.030 \\
$\omega$& ($\degr$)              & 260.53& 12.91 \\
{\em a}~$\sin$(i) & (${\rm 10}^{-3}$~au) & 0.256 & 0.022 \\
{\em K} & (km~${\rm s}^{-1}$) & 0.634 & 0.067 \\
{\em f}(m) & (${\rm 10}^{-7}$~${\rm M}_{\odot}$) & 0.868 & 0.227 \\
\hline
\end{tabular}
\end{minipage}
\end{table}

The spectra of $\epsilon$~UMa were retrieved from the ELODIE archive \citep{b13}.
This archive contains the complete collection of high-resolution
echelle spectra using the ELODIE fiber-fed echelle spectrograph \citep{b14}
mounted on the 1.93-m telescope at the Haute-Provence Observatory (France).
The spectra have a resolution ($\lambda$/$\triangle$$\lambda$) of about 42000.
The archived signal-to-noise ratio was between $\sim$200 and $\sim$400
in the spectral region near $\lambda$~5550~\AA.
In addition, eight spectra obtained 15 and 16 June 2000
at the same telescope were retrieved from the Hypercat Fits Archive
(http://leda.univ-lyon1.fr/11/spectrophotometry.html).
In Table~1, each spectrum is presented by its file name, exposure time,
Julian date of the observations and its corresponding phase.
Note that the phases were computed with the ephemeris from Table~2
(see below).

Special technique of radial velocity measurement of the hydrogen lines
was used and should be explained.
Well known, that in slowly rotating late B-type stars
the hydrogen lines have broad wings and sharp cores.
For $\epsilon$~UMa the value of $v$~sin~$i$ is equal 35~km~${\rm s}^{-1}$
\citep{b11}.
Unfortunately, the wings of the hydrogen lines can be affected by different
spectral lines.
For example, in spectrum of $\epsilon$~UMa the wing of $H_{\beta}$ line
affected by many Cr lines \citep{b15}.
On the other hand, the sharp cores of the hydrogen lines are not affected
by spectral lines and the flux is formed in the upper layers of
the atmosphere.
Thus our technique was to use only the sharp cores of the hydrogen lines
in spectra of $\epsilon$~UMa. The spectra were processed
using the spectral reduction software SPE developed by S. Sergeev at
Crimean Astrophysical Observatory (CrAO).
The program allows detecting the variations of the centre gravity of
the sharp cores of the hydrogen lines.
After processing, all the spectra were corrected for the motion of the Earth
around the Sun.
For example, the demonstration of the positional variability of the core
$H_{\alpha}$ line is presented in Fig.~1.
In Table~1, each spectrum is presented by its radial velocities computed from
the hydrogen lines, the mean radial velocity and errors of the mean
radial velocity.

\begin{figure}
\vspace{0.2cm}
\centerline{\includegraphics[width=70mm]{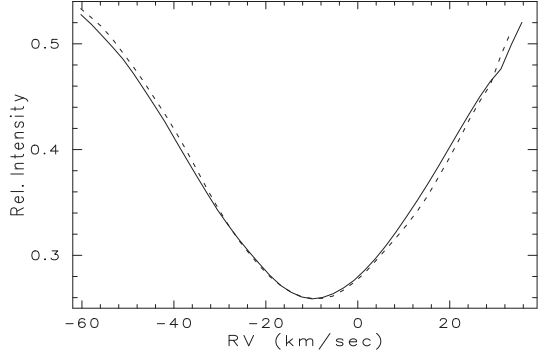}}
\caption{The observed intensity profiles of the cores of $H_{\alpha}$ line
 obtained on May 21, 2000 (phase = 0.93) and on June 6, 2000 (phase = 0.07)
 marked by the solid and dashed lines, respectively.
 The velocity scale is given with respect to the $\lambda$~=~6562.797~\AA.}
\label{figure1}
\end{figure}

\section{Orbital parameters}

Orbital elements have been determined by a non-linear least-squares fitting
of the mean radial velocities from Table~1 using the program BINARY
writing D.H. Gudehus
from Georgia State University (http://www.chara.gsu.edu/$\sim$gudehus/binary.html).
The solution for a single-lined binary is modelled by up to six parameters:
\begin{enumerate}
\item {\em P} period
\item {\em T} time of periastron passage
\item {\em e} eccentricity
\item ${V}_{\rm 0}$ system radial velocity
\item $\omega$ longitude of periastron
\item {\em a}~$\sin$(i) sine of inclination times semi-major axis
\end{enumerate}
The expected radial velocities are
\begin{equation}
RV = K[\cos{(\theta + \omega)} + e\cos{\omega}]
\end{equation}
where $\theta$ is the angular position of the star measured from the centre
of mass at a given instant.
The program also calculate the projected semi-amplitude variation of
the radial velocity:
\begin{equation}
 K = \frac{2 \pi a \sin{i}}{P \sqrt{1-e^{2}}},
\end{equation}
though this is never used as a parameter in the solution and
the mass function:
\begin{equation}
 f(m) = \frac{M^{3}_{2} {\sin}^{3}i}{(M_{1} + M_{2})^{2}}.
\end{equation}
The program BINARY gives the estimated standard deviations
of the orbital parameters as well.
The orbital solution of Table~2 was obtained by fitting a Keplerian
orbit to the 29 ELODIE radial velocity measurements.
Note that most of observations were obtained between
JD=2451568 (January 2000) and JD=2451714 (June 2000) (see Table~1).

Experience shows that the best Keplerian fit to the data with the fixed
period {\em P}~=~5.0887 d, the eccentricity {\em e}~=~0.503 and
the semi-amplitude {\em K}~=~0.634~km~${\rm s}^{-1}$.
The parameters of the best Keplerian orbital solution for $\epsilon$~UMa
are presented in Table~2.
In close binary system with Bp-Ap stars, there is evidence for a tendency
toward synchronization between the rotational and orbital motions.
This effect is thought to be produced by the tidal forces \citep{b21}.
The radial velocity curve is displayed in Fig.~2 with the residuals around
solution.
A linear trend is not observed in the residuals around the orbital solution
that can be explained by the absent of a second companion in
a longer-period orbit.
Although, the weighted r.m.s. around the best Keplerian solution
($\sigma$(O-C)) is equal to 0.131~km~${\rm s}^{-1}$.
This value is a bit large compared to the typical radial velocity
measurements from the ELODIE cross-correlation function \citep{b17}.

\begin{figure}
\vspace{0.25cm}
\centerline{\includegraphics[width=70mm]{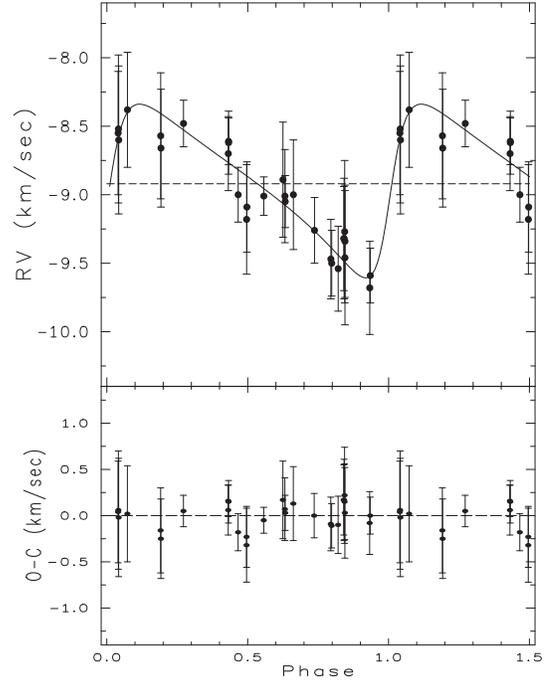}}
\caption{$Top:$ Phase diagram of the radial velocity measurements and
 Keplerian orbital solution for $\epsilon$~UMa.
 $Bottom:$ Residuals around the solution.}
\label{figure2}
\end{figure}

\section {Discussion}

The presence of the spots on the stellar surface of $\epsilon$~UMa
can change
the observed spectral line profiles and induces a periodic radial velocity
signal similar to the one expected from the presence of a satellite.
The hydrogen lines analysis is one of the best tools to discriminate
between radial velocity variations due to changes in the spectral line
shapes and variations due to the real orbital motion of the star.
It is obviously of interest to compare the phase diagram of the radial
velocity of $\epsilon$~UMa derived above with the phase diagram of
the radial velocity computed from metallic lines.
In this way, we selected the spectral region of the very prominent
unblended Cr~II line at $\lambda$~4558~\AA.

\begin{figure}
\vspace{0.15cm}
\centerline{\includegraphics[width=70mm]{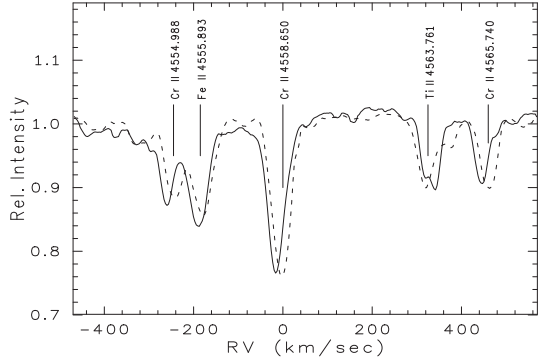}}
\caption{Positional variations of strong photosphere lines in the
 spectral region of the Cr~II line at $\lambda$~4558~\AA\
 obtained on June 15, 2000 (phase = 0.84) and on January 26, 2000
 (phase = 0.19) marked by the solid and dashed lines, respectively.
 The velocity scale is given with respect to the $\lambda$~=~4558.65~\AA.}
\label{figure3}
\end{figure}

Figure~3 shows the strong photosphere lines in the spectral region of
the Cr~II line at $\lambda$~4558~\AA\ obtained at phases before and after
the epoch of periastron passage. The radial velocity shift of the Cr~II
$\lambda\lambda$~4554, 4558 and 4565~\AA\AA\ lines and the Fe~II
$\lambda$~4555~\AA\ line at different phases is clearly seen.
The Ti~II line at $\lambda$~4563~\AA\ shows the line doubling that
appears at phase 0.84 and disappears at phase 0.19.
The maximum splitting of this line is at phase 0.04.
The splitting of the lines was first observed by \citet{b2} in the
spectrum of $\epsilon$~UMa.
However, the authors concluded that the doubling is not caused by
orbital motion but may be due to a combination of the physical effects
with Doppler effect in a rotating star.

\citet{b19} estimated the masses for 10 magnetic CP stars using the
position of stars in the log$R_{\sun}$~--~log$T_{\rm eff}$ plane.
Position of $\epsilon$~UMa in the log$R_{\sun}$~--~log$T_{\rm eff}$ plane
seems to be among the most evolved CP stars and
gives the value of $M_{\epsilon{\rm UMa}}$~=~3.0$\pm$0.4$M_{\sun}$.
According to \cite{b11} the radius of $\epsilon$~UMa is equal to
4.2$\pm$0.2$R_{\sun}$ corresponding to their choice of
$v\sin i$~=~35~km~${\rm s}^{\rm -1}$ and $i$~=~${\rm45}^{\circ}$ using
the trigonometric parallax measured by {\it Hipparcos}, $\pi$~=~40.30~mas
(ESO, 1997) and an angular diameter of 1.561~mas.
The effective temperature of $\epsilon$~UMa was taken from the paper
by \citet{b20} and is equal to 9340$\pm$530~K.
On the hypothesis that the rotational axis of $\epsilon$~UMa is perpendicular
to the orbital plane we can estimate the mass of the secondary star.
For the value of $M_{\rm 1}$~=~3.0$M_{\sun}$ and $i$~=~${\rm45}^{\circ}$
Eq.(3) gives $M_{\rm 2}$~=~0.014$M_{\sun}$.
This result gives a value $\sim$14.7$M_{\rm Jup}$, strongly suggesting that
the companion is in the typical brown-dwarf regime.

If we know the value $M_{\rm 1}$/$M_{\rm 2}$ then it is possible
to estimate the projected semi-amplitude variation of the radial velocity
for the companion according to formula:
\begin{equation}
 K_{2} = K_{1}\frac{M_{1}}{M_{2}},
\end{equation}
where $K_{1}$ is the projected semi-amplitude variation of the radial velocity
for $\epsilon$~UMa taken from Table~2.
For the value of $K_{\rm 1}$~=~0.634~km~${\rm s}^{\rm -1}$ Eq.(4) gives
$K_{\rm 2}$~=~135.9~km~${\rm s}^{\rm -1}$.
Thus, we can estimate the sine of inclination times semi-major axis using
Eq.(2). Computation gives the value of $a_{2}$$\sin$(i)~=~0.055~au.
These estimates show that the proposed brown-dwarf is quite close to
the surface of $\epsilon$~UMa at periastron. But, such close orbits are
not new. For example, the subgiant star HD~118203 have the planet with
eccentric orbit ($e$~=~0.31), the period of $P$~=~6.1335~d, and is
close to its parent star ($a$~=~0.06~au) \citep{b22}.

Another way to interpret the radial velocity variations is the radial
pulsation of $\epsilon$~UMa.
\citet{b23} are analysed observations of $\epsilon$~UMa obtained with
the star tracker on the {\it Wide Field Infrared Explorer} satellite.
The authors observed that a light curve has about 2 per cent amplitude of
photometric variability.
On the other hand, \citet{b24} has presented ultraviolet light curves
for $\epsilon$~UMa from the OAO-2 satellite which indicate that
the photometric variations of this star are due to variable ultraviolet
absorption effects which redistribute flux into the visible region
(see his Fig.~5). Note that no colour index changes \citep{b3}.
Certainly the radial pulsation appears unlikely given that the rotational
period is synchronized with the orbital period of $\epsilon$~UMa.

\section {Conclusions}

The archival ELODIE high-resolution echelle spectra of $\epsilon$~UMa
permit us to analyse the radial velocity variations of the sharp cores
of the hydrogen lines.
This allowed determining the orbital elements of binary system for
the CP star $\epsilon$~UMa.
The best Keplerian fit to the data shown that the rotational period
is synchronized with the orbital period.
We are estimated the mass of the secondary star which is equal
$\sim$14.7$M_{\rm Jup}$. This result indicate that the companion is
the brown-dwarf with the projected semi-amplitude variation of the
radial velocity $K_{\rm 2}$~=~135.9~km~${\rm s}^{\rm -1}$ and
the sine of inclination times semi-major axis
$a_{2}$$\sin$(i)~=~0.055~au.

\section*{Acknowledgements}
The author would like to thank the referee Dr. J.B. Rice of this Letter
for his extremely helpful comments.

\bsp

\label{lastpage}

\end{document}